\documentclass{emulateapj}

\shorttitle{Submillimeter observations of Distant Red Galaxies}
\shortauthors{Knudsen et al.}

\begin{document}


\title{Submillimeter observations of Distant Red Galaxies:  uncovering the 1\,mJy 850$\mu$m-population}


\author{K.~K.\ Knudsen\altaffilmark{1}, 
P.\ van der Werf\altaffilmark{2}, 
M.\ Franx\altaffilmark{2}, 
N.~M.\ F\"orster Schreiber\altaffilmark{3}, 
P.~G.\ van Dokkum\altaffilmark{4},
G.D.~Illingworth\altaffilmark{5},
I.\ Labb\'e\altaffilmark{6}, 
A.~Moorwood\altaffilmark{7}, 
H.-W.\ Rix\altaffilmark{1}, 
G.\ Rudnick\altaffilmark{8}}

\altaffiltext{1}{Max-Planck-Institut f\"ur Astronomie, K\"onigstuhl 17, D-69117 Heidelberg, Germany; knudsen@mpia-hd.mpg.de}
\altaffiltext{2}{Leiden Observatory, P.O.Box 9513, NL-2300 RA Leiden, The Netherlands}
\altaffiltext{3}{Max-Planck-Institut f\"ur extraterrestrische Physik, Giessenbachstrasse, Postfach 1312, D-85748 Garching, Germany} 
\altaffiltext{4}{Department of Astronomy, Yale University, P.O.\ Box 208101, New Haven, CT 06520-8101}
\altaffiltext{5}{UCO/Lick Observatory \& Department of Astronomy, University of California, Santa Cruz, CA 95064}
\altaffiltext{6}{Carnegie Observatories, 813 Santa Barbara Street, Pasadena, CA 91101}
\altaffiltext{7}{European Southern Observatory, Karl-Schwarzschild-Strasse 2, D-85748 Garching, Germany}
\altaffiltext{8}{National Optical Astronomy Observatory, 950 North Cherry Avenue, Tucson, AZ 85719}



\begin{abstract}
We present a study of the submillimeter (submm) emission of 
Distant Red Galaxies (DRGs).
The DRGs are selected by the criterion $J-K > 2.3$, and are
generally massive galaxies at redshifts higher than 2, with red
rest-frame optical colors.
Using a deep SCUBA submm image of a field centred on the
cluster MS\,$1054{-}03$, we obtain a statistical detection of the DRGs 
at redshift $z=2-3.5$, with
an average 850\,$\mu$m flux density of $1.11\pm0.28$\,mJy. 
The detection implies an average star formation rate (SFR) of 
$127\pm34$\,M$_\odot$\,yr$^{-1}$ (lensing corrected),
assuming that the far-infrared (FIR) spectral energy distribution (SED) 
is well-described by a modified blackbody. 
The SFR derived from the submm agrees well with 
SFRs derived from SED fitting 
of optical-near-infrared data and average X-ray emission.
Constant Star Formation models imply ages of 2Gyr, extinction 
$A_V=2.4$\,mag, which is consistent with the FIR to rest-frame optical luminosity ratio of $\sim 15$.
DRGs are older and have lower SFRs relative to 
optical luminosity than (ultra-)luminous infrared galaxies, 
although their FIR luminosities are similar. 
The DRGs at $2<z<3.5$ and the Extremely Red Objects ($I-K>4$) at $1<z<2$, 
which were also investigated, contribute
5.7 and 5.9 Jy deg$^{-2}$ respectively to the submm background. 
Simple estimates suggest that 
these populations contribute
$\sim50\%$ of the flux from sources with $0.5<f_{850}<5$\,mJy, 
which is where the peak of energy is produced. 
We have therefore uncovered one of the most important populations of galaxies
contributing to the sub-mm background.

\end{abstract}






\section{INTRODUCTION}

Members of the recently defined galaxy population {\it Distant Red Galaxies} 
(DRGs) 
are identified by their red observed near-infrared colors, $J-K>2.3$
(\citealt{franx03,vDokkum03,forster04} (henceforth FS04)).
Galaxies selected by this color cut are generally at $z\sim 2-3.5$, 
and are hence selected on their red colors in the rest-frame optical. 
Their stellar populations are relatively old, or are highly dust obscured. 
DRGs account for over half of the cosmic stellar mass at the high mass
end at redshifts 
$2\leq z \leq 3.5$, somewhat more than the contribution 
from Lyman Break Galaxies (LBGs) (\citealt{rudnick03}; FS04).  
One of the most pressing questions concerning these galaxies is
how high their star formation rates (SFRs) are. 
Based on  near-infrared (near-IR) spectroscopy, high star formation 
rates of $200-400\,M_\odot$\,yr$^{-1}$ and high stellar masses of 
(1$-$5)$\times$10$^{11}$\,M$_\odot$ were determined \citep{vDokkum04}. 
Unfortunately, near-IR spectroscopy is available for only a few
galaxies at this moment.
Star formation rates can also be estimated using other techniques,
especially
using radio, X-ray or submillimeter (submm) observations.
Most of the DRGs are too faint to be detected individually 
at these wavelengths, but it
is possible to estimate their average star formation rate by 
``stacking'' the fluxes for a large number of sources.
Using this technique, \citet{rubin04} have detected the X-ray emission of
DRGs in the field of the cluster MS\,$1054{-}03$ 
and derived an average star formation
rate of $214\pm99$\,M$_\odot$\,yr$^{-1}$ 
if all the X-ray flux is attributed to the 
SFR.  However, the flux contribution from faint AGNs remains 
unclear.

Here we present a different SFR measurement for the DRGs, 
based on the average 850$\,\mu$m flux density, which  
we attribute to emission from dust that has been heated by young stars.
We have taken very deep SCUBA $850\,\mu$m observations of the field of
the cluster MS\,$1054{-}03$
(\citealt{knudsenphd}; Knudsen et al.\ {\it in prep}). 
These deep data provide a unique opportunity for 
studying the submm properties of DRGs.  
Throughout this Letter we will assume an $\Omega=0.3$, $\Lambda=0.7$
cosmology with $H_0=70$\,km\,s$^{-1}$\,Mpc$^{-1}$. 
Unless otherwise stated, we use Vega magnitudes. 


\section{SUBMILLIMETER DATA}

Submillimeter data for MS\,$1054{-}03$ were obtained using the 
Submillimetre Common-User Bolometer Array (SCUBA; \citealt{holland}) 
at 850\,$\mu$m on the 15\,m James Clerk Maxwell Telescope at Mauna Kea, 
Hawaii. Total integration time was 49\,hours, distributed almost 
evenly over 3
pointing centers to cover the entire field of interest.
The data were reduced using SURF \citep{surf}, and sources were
extracted using 
the Mexican Hat Wavelet algorithm \citep{barnard04,knudsen05}. Noise
properties of the data were determined using Monte Carlo 
simulations of artificial datasets. 
Details of these reduction and analysis methods are described in 
\citet{knudsenphd} and \citet{knudsen05}.
The total area of the map is 14.4\,arcmin$^2$, though for 
the analysis the outer 23$''$ edge has been trimmed off due to high 
noise leaving a useful area is 12.8\,arcmin$^2$.  
The image has an area-weighted $1\sigma$ r.m.s.\ noise of $\sim
1.2$\,mJy\,beam$^{-1}$, with variations between 0.9 and
1.6\,mJy\,beam$^{-1}$.
The flux calibration uncertainty at 850\,$\mu$m is about 10\%. 
The Full Width Half Maximum (FWHM) of a point source in the map is 
15$''$. The astrometric accuracy is approximately $3''$, 
due to the pointing uncertainty of the JCMT.

Nine submm sources are detected with flux densities of $3.5-5.0\,$mJy 
with signal-to-noise ratio $>3$.  
Two  of the nine submm sources have been securely identified, both 
with DRGs, namely the galaxies MS-1383 which has a spectroscopic 
redshift of 2.423 \citep{vDokkum04} and MS-723 which has a photometric 
redshift of 1.88$^{+0.12}_{-0.02}$ \citep{forster05}.
Five of the other submm sources have likely identifications, 
with at least one DRG or Extremely Red Object (ERO, defined by
an observed color $I-K>4$)
nearby.   The two remaining sources have no 
obvious counterpart.  
A mass model for the foreground cluster \citep{hoekstra00}  
shows that the gravitational lensing magnification is small 
for the sources, on average around 20\%.


\section{SUBMM EMISSION FROM DISTANT RED GALAXIES}
\label{sect:statan}

\subsection{Statistical analysis}

We measured the characteristic submm fluxes for DRGs as a class 
of galaxies by measuring the flux in the 850\,$\mu$m map at the 
position of all  the galaxies in the sample.  
We found an average 850\,$\mu$m flux of $1.19\pm0.22$\,mJy for 30 DRGs 
with $K < 22.5$, and with an uncertainty on their $K$-band magnitude 
smaller than 0.1, and minimum weight in the photometry $> 0.2$ 
(this measure is related to the total integration time -- see FS04), 
present within the area covered by the SCUBA map. 
As the sensitivity is not 
uniform across the field, we determine the weighted mean and 
standard deviation, weighted
by the squared reciprocal of the 1$\sigma$ r.m.s.\ noise at the position.  
The weighted 
mean compared to the unweighted mean only changes the result by a 
few percent, indicating that the weighted result is not dominated by a 
few data points with high weights.
\citet{webb_ero} discussed a possible bias in this procedure due to 
confusion.  If two or more DRGs are separated by less than 
the 850\,$\mu$m beam, emission might be counted twice.  This is 
corrected for by estimating the flux contribution from each of the 
close-by DRGs to the flux measured at the given position 
(for details see: \citealt{webb_ero}).  
We find that this effect influences our measurements
by $\sim$\,2$-$10\%.  

\begin{figure}[!t]
\epsscale{.70}
\plotone{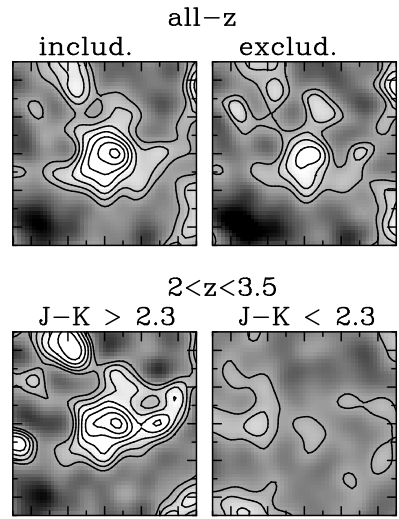}
\caption[]{
Postage stamp sized sub-images from the statistical stacking analysis 
of the DRGs in the field MS1054$-$03. 
Each sub-image is $50''\times 50''$. 
The images represent a weighted mean.  The overlayed contours represent 
a mean 850\,$\mu$m flux of 0.0,0.2,..,1.2\,mJy.  
Top row:  DRGs for all redshifts, including (left) and excluding (right) 
sources associated with the discrete SCUBA sources. 
Bottom row: sources at redshift $2-3.5$, left are DRGs, right non-DRGs.  
\label{fig:stack_res}}
\end{figure}

As mentioned above, two DRGs are identified as the counterparts
of two individual SCUBA detections.  
Additionally, three other DRGs are found within
the identification search radius of three other SCUBA detections, 
though these are not secure identifications.
We calculate the mean 850\,$\mu$m flux density, both including and
excluding these five objects.
Results are summarized in Table \ref{tab:stack}.
If we omit the DRGs associated with discrete SCUBA sources, 
we obtain an average flux of $0.74\pm0.24$\,mJy, 
demonstrating that the detection is not caused solely by these galaxies. 
We show the average images of both samples in Fig.~\ref{fig:stack_res}. 
The emission is well-centered.

\begin{deluxetable*}{lccccccccccc}[ht]
\tablecolumns{12}
\tabletypesize{\footnotesize}
\tablewidth{0pc}
\tablecaption{The mean 850\,$\mu$m flux of DRGs \tablenotemark{1} \label{tab:stack}}
\tablehead{
\colhead{} & \colhead{} & \multicolumn{6}{c}{Including all}  & \colhead{} & \multicolumn{3}{c}{Excluding discrete sources} \\
\cline{3-8} \cline{10-12} \\
\colhead{Sample}  & \colhead{$K_{limit}$} & \colhead{$N$} & \colhead{$\langle f_{850}\rangle$} & \colhead{\% EBL} & \colhead{$\langle {\rm SFR}\rangle$} & \colhead{$\langle L_V\rangle$} & \colhead{$\langle {\rm SFR}\rangle / \langle L_V \rangle$} && \colhead{$N$} & \colhead{$\langle f_{850}\rangle$} & \colhead{$\langle {\rm SFR}\rangle /\langle L_V\rangle$}  \\
\colhead{(1)} & 
\colhead{(2)} & 
\colhead{(3)} & 
\colhead{(4)} & 
\colhead{(5)} & 
\colhead{(6)} & 
\colhead{(7)} & 
\colhead{(8)} & &
\colhead{(9)} & 
\colhead{(10)} &
\colhead{(11)} \\
}
\startdata
all-$z$            & 22.5 & 30 & $1.19\pm0.22$ & 21.8(30.9) & 169 & 7.3 & 23.2 && 25 & $0.74\pm0.24$  & 15.1 \\
2$<$z$<$3.5        & 22.5 & 18 & $1.11\pm0.28$ & 13.0(18.4) & 159 & 7.9 & 20.2 && 15 & $0.69\pm0.30$  & 13.8 \\ [1.3ex]
\raisebox{1.0ex}[0pt]{no-DRGs} & 22.5 & 71 & $0.14\pm0.15$ &  \phn6.5\phn(9.3)  &  \phn20 & 4.7 &  \phn4.3 && 67 & $0.04\pm0.15$ & \phn1.2  \\ 
\raisebox{1.5ex}[0pt]{2$<$z$<$3.5}&    &               &            &     &     &      &&    &                &      \\ [0.2ex]
CSF                & 21.7 &  \phn9 & $1.67\pm0.38$ & 10.3(14.6) & 238 &10.4\phn & 22.9 &&  \phn7 & $1.15\pm0.43$  & 17.7 \\ 
SSP                & 21.7 &  \phn5 & $0.67\pm0.54$ &  \phn1.6\phn(2.3)  &  \phn96 & 6.1 & 15.7 && \nodata & \nodata  & \nodata 
\enddata
\tablenotetext{1}{Uncorrected for the average gravitational lensing of 20\%.}
\tablenotetext{\mbox{}}{Units:  (2) [mag]; (4) [mJy]; (5) \% of the EBL value from \citet{fixsen98}(\citet{puget96}); (6) [M$_\odot$/yr] ; } 
\tablenotetext{\mbox{}}{(7) [$10^{10}$\,L$_\odot$]; (8) [M$_\odot$/yr/$10^{10}$\,L$_\odot$];  (10) [mJy]; (11) [M$_\odot$/yr/$10^{10}$\,L$_\odot$] .}
\end{deluxetable*}

For DRGs restricted to the redshift interval 2\,$<$\,$z$\,$<$\,3.5, 
we obtain an average flux of $1.11\pm0.28$\,mJy for the DRGs.
For comparison, the average flux for galaxies 
at 2\,$<$\,$z$\,$<$\,3.5 which are not DRGs, i.e.\ $J-K <2.3$ 
(henceforth refered to as {\em non-DRGs}), is substantially
lower at only $0.14\pm0.15$. 

Following FS04, 
we fitted the full optical/near-IR spectral energy distribution (SED)
with stellar population models for DRGs with $K$\,$<$\,21.7 
and 2\,$<$\,$z$\,$<$\,3.5.  
The sample was separated into 
those best fit by constant star formation models (CSF) or single
age stellar populations (SSP).
The CSF-subsample has an average flux of $1.67\pm0.38$\,mJy, 
while the SSP-subsample has an average flux of $0.67\pm0.54$.
Excluding the DRGs associated with discrete SCUBA sources 
affects only the CSF subsample and its mean flux decreased 
to $1.15\pm0.43$\,mJy.  
We note that the lensing correction is less than or comparable 
to the uncertainty of the average fluxes.

\subsection{Monte Carlo simulations}

To test the reliability of the detection of the mean 850\,$\mu$m 
flux from the  DRGs, Monte Carlo (MC) simulations were performed 
by making the statistical measurements
at random positions in the field.  
The MC simulations were repeated 10000 times. 
As with the real data, the MC measurements 
were corrected for the confusion of two or more sources separated 
by less than the distance of a beam. 
For 45-50 random positions on the map, 
the weighted mean is 0.13$\pm$0.17\,mJy.
As the noise distribution deviates from Gaussian, the probability 
for getting a $>$\,3$\sigma$ detection at random positions is 2.4\,\%. 
The probabilities for a $>$\,4$\sigma$ detection and a $>$\,5$\sigma$ 
detection are 0.2\,\% and 0.02\,\% respectively. 
A similar result is seen for 20-30 
and for 10-20 random positions.   Thus we conclude that the detected 
mean 850\,$\mu$m flux  is real. 

\subsection{Possible systematic uncertainties}

We performed simulations to verify how reliable the average
detections were. We assumed that the star formation rate is
proportional
to restframe $L_V$ luminosity, with different ratios for
different classes of objects. We found that systematic errors
occur at the level of 0.1-0.2 mJy for the average fluxes. For the
DRGs and EROs this is not a concern, but
for the fainter objects (e.g., the non-DRGs
with average flux 0.14 mJy) this  implies that the systematic errors
are at least as large as the detections.
The systematic problems are  due to a combination of effects,
including the negative ``sidelobes'' in the beam pattern, which
cause the total flux of the map to be zero, and can reduce the
flux for the faint, but abundant non-DRGs, and superpositions  of
objects. 
Obviously, similar observations of other fields would help to
constrain
the errors for the DRGs; but it is unclear whether such measurements will
ever constrain the flux from the faint non-DRGs.


\section{DISCUSSION}

\subsection{Implied star formation rates of the DRGs}

The average SFR of the DRGs can be estimated 
from the average submm flux assuming an  
SED and an initial mass function (IMF). 
As the submm and far-infrared (FIR) 
SED of the DRGs is unknown, we use the SED description for 
dusty starburst galaxies from \citet{YunCar02} for calculating 
the FIR luminosity. 
Because of the large negative $k$-correction in the submm, the observed 
850\,$\mu$m flux density is essentially constant between redshift 1 and 
8 for a given luminosity. 
For the DRG 2\,$<$\,$z$\,$<$\,3.5 sample, the average FIR luminosity 
is $1.2\times10^{12}$\,L$_\odot$, which is comparable to the 
luminosity of the local ultraluminous infrared galaxy (ULIRG) Arp\,220.   
Furthermore, we base the conversion between FIR luminosity 
and SFR on  \citet{kennicutt98}, where a Salpeter IMF in the mass range 
0.1-100\,M$_\odot$ is assumed.
However, we assumed a mean age of the stellar population of 1Gyr 
and a constant star formation during that time,
which lowers the conversion by a factor of 1.5: 
\mbox{{\em SFR}[M$_\odot$\,yr$^{-1}$]$= 1.17\times 10^{-10} L_{FIR}$\,[L$_\odot$]}.
We assumed this longer mean age since modeling of the optical-Near IR
SEDs suggest ages of 1-2 Gyr (FS04).
We obtain an average SFR of $159\pm42$\,M$_\odot$\,yr$^{-1}$ 
for $2<z<3.5$, 
where the error reflects the accuracy of the submm detection. 
When correcting for gravitational lensing, this is an $\langle$SFR$\rangle$
of $127\pm34$\,M$_\odot$\,yr$^{-1}$.  
We note that the FIR luminosity and SFR have considerable
uncertainties as they depend on the assumed
shape of the SED and  IMF. 

The average SFR derived from the thermal dust emission in the 
far-IR ($\lambda_{rest}$\,$\sim$\,250\,$\mu$m) can be compared to the SFRs 
inferred from the SED fits to the optical and near-IR data (FS04).
The SFRs from SED fits are rather model-dependent.
One of the main uncertainties is the parametrization of the star
formation
history (constant, declining etc). The maximum is generally given by
constant formation rates, and is on average $\sim$170\,M$_\odot$\,yr$^{-1}$
for the sample studied here (FS04). 
This compares well with the SFR derived from the submm flux.

The good agreement suggests that the majority of DRGs are well approximated 
by simple models with constant star formation rate, ages of 2 Gyr,
and high extinction $A_V=2.4$ (FS04). 
Despite the abundant dust 
the overall ratio of $L_{FIR}/L_{opt}\sim 15$, which is in good agreement 
with $A_V$, 
is lower than that of ULIRGs in the nearby universe. 
This implies that DRGs are forming stars more steadily 
over longer periodes than nearby starbursts. 
We expect to find very large
reservoirs of cold gas to sustain these high star formation rates.

When interpreting the average rates it should be noted that 
not all DRGs are likely to be forming stars. In a deep
Spitzer study of DRGs in HDF-South, \citet{labbe05} found that 30\% are
best fit by old stellar populations at redshifts higher than 2,
and the remaining are best fit by constant star formation models.
The results suggest that at redshifts higher than 2, the majority of
massive galaxies were still forming stars at rather high rates, enough
to build up their stellar mass in 1-2 Gyr.

\subsection{Contribution to the comoving star formation density}

We derive a comoving star formation density of about 
0.045\,M$_\odot$\,yr$^{-1}$\,Mpc$^{-3}$ for the DRGs with $2<z<3.5$.
This can be compared with the 
comoving star formation density
derived from UV selected galaxies at $z=2-3.5$ of  approximately
0.022 to 0.11 \,M$_\odot$\,yr$^{-1}$\,Mpc$^{-3}$ (Steidel et al 1999),
excluding, and including a correction for dust absorption.
As the dust correction is uncertain for UV selected galaxies,
especially
at the faint end which contributes a large fraction of the flux, we
have to conclude that the contributions of the two populations
are similar, but still uncertain. The main uncertainty for the DRGs 
is the luminosity function at the faint end, and the main
uncertainty for the LBGs is the dust correction.

We note that several authors have tried to detect the  average submm
flux of the LBGs (e.g., Chapman et al. 2000, Webb et al 2003),
without statistically significant results. Since our simulations
suggest
that systematic effects can be large at the very faint end, 
the average 850\,$\mu$m flux of these sources remains an open question. 

\subsection{Dissecting the contribution to the sub-mm background}

Each population of high redshift galaxies contributes to the submm 
Extragalactic Background Light (EBL). 
Previous studies have addressed this for the 
LBG and the ERO populations.  
There has been no significant statistical detection of the 
submm emission from LBGs. 
Through SCUBA 
observations, \citet{chapman_lbg} placed upper limits on the LBG 
contribution to the submm EBL of 0.2\%, while \citet{webb_lbg} placed 
an upper limit of 20\% when extrapolating to a redshift interval 
1\,$<$\,$z$\,$<$\,5. 
Contrary to the LBG case, statistical detection of the ERO population 
has been successful.  \citet{wehner} and \citet{webb_ero} find 
an average submm flux of $1.58\pm0.13$\,mJy and $0.40\pm0.07$\,mJy
respectively, and determine the contributions to the submm EBL 
to be about half and 10\%.  The discrepancy between the results 
can be primarily assigned to the difference in depth of the ERO 
selection and the strong clustering of EROs, as the fields studied 
by \citet{wehner} appear to have a large overdensity of EROs. 

Using the present study we constrain the contribution 
of the DRGs to the submm EBL. 
For the whole sample of DRGs in the MS\,1054$-$03 field, the total 
flux density is 
approximately 9.6\,Jy/deg$^2$, which is 21.8\% of the EBL 
value measured by \citet{fixsen98} of 44\,Jy/deg$^2$ and 30.9\% of
the value from \citet{puget96} of 31\,Jy/deg$^2$.
For the DRG sample limited to the redshift range $2<z<3.5$ the 
contribution is 5.7\,Jy/deg$^2$, i.e.\ 12.9\% [18.4\%] of the result from 
\citet{fixsen98} [\citet{puget96}]. 
We can perform the same measurement for the EROs in the interval 
$1 < z < 2$.  We apply similar magnitude and uncertainty cuts as 
for the DRGs (see section \ref{sect:statan}). 
We obtain an average flux of $0.69\pm0.21$\,mJy, by averaging 35 sources, 
which implies a comoving star formation rate density of 
0.085\,M$_\odot$\,yr$^{-1}$\,Mpc$^{-3}$.
The total contribution from the EROs is 5.9\,Jy/deg$^2$. 
Hence the total contribution to the submm EBL of the EROs at $1<z<2$ 
and the DRGs at $2<z<3.5$ is 11.6\,Jy/deg$^2$.

The surface density of the two populations combined is 
$\sim5$\,arcmin$^{-2}$ down to $K=22.5$\,mag, similar to the surface
density derived from the submm number counts, where 
$N(>\!0.5{\rm mJy})\sim 6.5$\,arcmin$^{-2}$.

We can use our measured ratios of SFR/$L_V$ to predict the fluxes of
the EROs and DRGs at $1 < z < 4$: 
we assume they all have the same ratio, i.e.,
SFR scales linearly with $L_V$. We assume that the ratio is that of
non-DRGs for the other galaxies at the same redshift interval.
Using this very simple assumption, we find that the EROs+DRGs
contribute 12.3 Jy deg$^{-2}$.
This is about 50\% the background between $0.5 <  f < 5$\,mJy determined
by \citet{knudsenphd}, which is 26\,Jy\,deg$^{-2}$.
For the non-EROs/DRGs at $ 1 < z < 4$ and $K < 22.5$ we estimate 1.4 Jy deg$^{-2}$.

It is interesting to speculate whether there exists an evolutionary 
link between the massive high redshift galaxy populations, namely the 
DRGs and EROs and the submm galaxies (SMGs) with $f_{850} > 5$\,mJy, 
which are found to have comparable masses \citep{greve05}.  
Possibly the SMGs represent a short-lived phase with a duration of
$\leq100$\,Myr \citep{chapman05}, which is $>10$ times shorter than 
what we find for the DRGs.  The surface density of bright SMGs is 
about 0.2\,arcmin$^{-2}$.  This relative to the surface density of 
DRGs and EROs, is comparable to the ratio of the duration of the 
phases. 

In summary, 
the submm detection of the DRGs strongly supports
that these high-$z$, massive galaxies are undergoing 
star formation at high rates. 
Furthermore, 
the results presented here, based on simple assumptions, 
imply that DRGs and EROs dominate
the counts around 1\,mJy.

\acknowledgments

We thank Tracy Webb for helpful discussions. 
The James Clerk Maxwell Telescope is operated by The Joint Astronomy Centre 
on behalf of the Particle Physics and Astronomy Research Council of the 
United Kingdom, the Netherlands Organisation for Scientific Research, and 
the National Research Council of Canada.
We thank the staff at the JCMT for their assistence during the
observations. 
Support from NSF CAREER grant AST-0449678 is acknowledged.

\clearpage 

\end{document}